\documentclass[conference]{IEEEtran}
\IEEEoverridecommandlockouts

\def\re{\,{\rm Re}}
\def\im{\,{\rm Im}}

\def\bmat#1{\left[ \begin{array}{#1}}
\def\emat{\end{array} \right]}
\def\bdet#1{\left| \begin{array}{#1}}
\def\edet{\end{array} \right|}

\def\bbox#1{\textbf{#1}}

\def\cxE{\bbox{E}}

\def\cxG{\bbox{G}}

\def\cxS{\bbox{S}}

\def\cxi{\bbox{i}}

\def\cxv{\bbox{v}}

\def\cxx{\bbox{x}}

\def\cxZ{\bbox{Z}}
\DeclareFontFamily{U}{omega}{}
\DeclareFontShape{U}{omega}{m}{n}{<5->omsegr}{}%
\DeclareFontShape{U}{omega}{b}{n}{<5->omsegrb}{}%
\DeclareFontShape{U}{omega}{m}{it}{<5->omsegri}{}%
\DeclareFontShape{U}{omega}{b}{it}{<5->omsegrbi}{}%
\DeclareSymbolFont{Omeb}{U}{omega}{b}{n}
\DeclareMathSymbol{\cxpsi}{\mathord}{Omeb}{"79}
\DeclareMathSymbol{\cxPsi}{\mathord}{Omeb}{"59}

\def\h1{\hspace{10mm}}

\usepackage{amsmath,amssymb}
\usepackage{float}
\usepackage{lscape}

\usepackage{microtype}
\usepackage{siunitx}
\usepackage{algorithm,algpseudocode}

\usepackage{multicol,multirow}
\usepackage{amsthm}
\usepackage{xparse}
\usepackage{graphicx}
\usepackage{tikz}
\usepackage{threeparttable}
\usepackage{comment}
\usepackage{makecell}
\usepackage{colortbl}
\def\BibTeX{{\rm B\kern-.05em{\sc i\kern-.025em b}\kern-.08em
		T\kern-.1667em\lower.7ex\hbox{E}\kern-.125emX}}

\usepackage{tabularx}
\usepackage{pgfplots}
\usetikzlibrary{calc,patterns,angles,quotes}
\usepackage{mathtools}
\usepackage{xfrac}
\usepackage{bm}
\usepackage{cite}

\pgfplotsset{compat=1.18}

\graphicspath{{./figures/}}

\graphicspath{{./figures/}}

\newcounter{problemC}

\newcommand{\ab}{\alpha\beta}

\newcommand{\dq}{\mathrm{dq}}
\newcommand{\dcom}{\mathrm{d}}
\newcommand{\qcom}{\mathrm{q}}

\newcommand{\pu}{\mathrm{pu}}

\newcommand{\rf}{\mathrm{ref}}

\newcommand{\ext}{\mathrm{set}}
\newcommand{\gr}{\mathrm{g}}

\usepackage{subcaption}
\allowdisplaybreaks

\DeclareSIUnit \pu {pu}
\DeclareSIUnit \voltampere {VA}

\pgfplotsset{compat=1.18}

\begin{document}

\title{On the input admittance of a universal power synchronization controller with droop controllers
}
%

\author{Orcun Karaca, Irina Subotic, Lennart Harnefors, Ioannis Tsoumas
	\thanks{O. Karaca is with ABB Corporate Research, Switzerland. email: {\tt orcun.karaca@ch.abb.com}. I. Subotic and I. Tsoumas are with ABB System Drives, Switzerland. emails: {\tt \{irina.subotic, ioannis.tsoumas\}@ch.abb.com}. L. Harnefors is with ABB Corporate Research, Sweden. email: {\tt lennart.harnefors@se.abb.com.} }}

\maketitle

\begin{abstract}
Recent work has proposed a universal framework that integrates the well-established power synchronization control into vector current control. Using this controller for the parallel operation of grid-forming converters, and/or with loads that have strong voltage magnitude sensitivity, requires additional loops manipulating the voltage magnitude, e.g., $QV$ and $PV$ voltage-power droop controllers. This paper derives the input admittance of the resulting overall scheme. Sensitivity analyses based on the passivity index demonstrate the benefits of the proportional components of $QV$ and $PV$ control. A case study is presented where a grid-forming converter is operated in parallel with a generator.\looseness=-1
\end{abstract}

\begin{IEEEkeywords}
	grid-forming control, power synchronization, passivity index
\end{IEEEkeywords}

\section{Introduction}
With the increasing penetration of renewable energy sources and the growing prominence of power electronic interfaces in power systems, the demand for grid-forming control methods is rising. These methods are characterized by their ability to emulate inertia, share power, and form voltage. Among the most extensively analyzed ones in both literature and industry are  droop control~\cite{chandorkar2002control,Yao2017}, power synchronization control (PSC)~\cite{zhang2009power}, virtual synchronous machine control~\cite{VSM2015}, virtual oscillator control~\cite{VOC2014}, synchronous power control~\cite{zhang2016synchronous} and grid-forming vector current control~\cite{Schweizer2022}. Recently, the authors in~\cite{harnefors2020universal} integrated the well-established PSC into vector current control that has inherent current limiting capability. This framework offers greater flexibility with a point-of-common-coupling (PCC) voltage alignment of the references. Additionally, it ensures a smooth transition from more conventional control schemes, e.g., those using a phase-locked loop (PLL). \looseness=-1

The parallel operation of grid-forming converters and their operation with loads that have strong voltage magnitude sensitivity require the addition of voltage control loops. A $QV$ control functionality is essential for reactive power sharing during the parallel operation of converters and/or synchronous generators~\cite{chandorkar2002control}. For example, during black starts, a $QV$ droop is sufficient to synchronize the magnitude ramps of different converters. On the other hand, a $PV$ control functionality is needed to push power over loads with strong voltage magnitude sensitivity, such as diode rectifiers~\cite{yu2019analysis,karaca2025effective}, and it is desirable also in relatively resistive microgrid settings~\cite{guerrero2006wireless}. To this end, this paper presents a universal power synchronization controller (UPSC) with $QV$ and $PV$ droop controllers. \looseness=-1

Deriving the small-signal input admittance of a grid-connected converter and employing passivity or Nyquist criteria are common practices to ensure stability, optimally tune controllers, and obtain valuable insights into control behavior~\cite{harnefors2007input,harnefors2007modeling,harnefors2015passivity,wen2015analysis,zhao2022low,chen2024limitations,chen2024extended}. Similarly, modeling the input admittance of the UPSC with $QV$ and $PV$ droop controllers holds great value as it would allow us to better understand the impact of different control gains. To the best of our knowledge, its derivation and analysis are not available in the literature. \looseness=-1

Our contributions are as follows. We derive the input admittance of UPSC with $QV$ and $PV$ droop controllers. Using the passivity index, we illustrate the benefits of the proportional components of these controllers, i.e., $QV$ and $PV$ droops. 
A large integral gain in $PV$ control instead creates a passivity dip, resulting in performance limitations in certain applications. Finally, a case study based on the operation of a grid-forming converter in parallel with a generator confirms insights obtained from the passivity index when droop gains are reduced and voltage integral gain is increased. \looseness=-1

\section{Preliminaries}

\textit{Notation:} Adopting the notation from \cite{harnefors2007modeling}, complex space vectors in the (synchronous) $\dq$-frame are denoted by boldface letters,~$\cxi$. 
When they refer to the (stationary) $\ab$-frame, the superscript~$^s$ is added,~$\cxi^s$. 
Transfer functions that map between complex space vector spaces are denoted by $\cxG(s)$. Here, the Laplace variable $s$ is to be interpreted as the derivative operator $s=\frac{d}{dt}$, where appropriate. If these functions are real, implying no cross-coupling, then italic letters are used instead; $G(s)$. 

Complex-coefficient functions are applicable only to $\dq$-frame symmetric systems. The $\dq$-frame asymmetry requires a MIMO function matrix denoted by slanted boldface letters,\looseness=-1
\begin{equation}
	\bm{G}(s) = \begin{bmatrix}
		G_{\dcom\dcom}(s)	& -G_{\dcom\qcom}(s)\\ G_{\qcom\dcom}(s)& G_{\qcom\qcom}(s)
	\end{bmatrix}.
\end{equation}
The function above maps between real space vectors,~$\bm{i}$. Whenever it is convenient to do so, we switch between complex space vectors and their real space vector equivalents,
$\bm{i}=\begin{bmatrix}
		i_{\dcom} &
		i_{\qcom}
	\end{bmatrix}^\top=\begin{bmatrix}
		\re\{\cxi\} &
		\im\{\cxi\}
	\end{bmatrix}^\top,\, \text{or}\ \cxi=i_{\dcom}+ji_{\qcom}.$
The superscript~$^*$ denotes the (Hermitian) conjugate, whereas the superscript~$^c$ denotes the quantities in the converter $\dq$-frame. In the remainder, all quantities (including frequency and time) are normalized. A first-order low pass filter is denoted by $H_{\alpha}(s)=\frac{\alpha}{s+\alpha}$, and $\alpha$ is referred to as its bandwidth. 

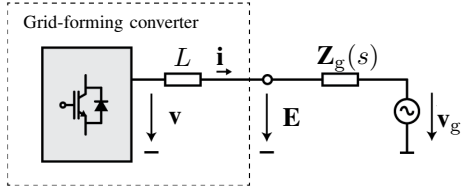
\begin{figure}[t]
	\centering
	\centering
	{\begin{tikzpicture}[every text node part/.style={align=center}]
			\node[draw=none,fill=none] at (0,0){\includegraphics[width=0.65\columnwidth]{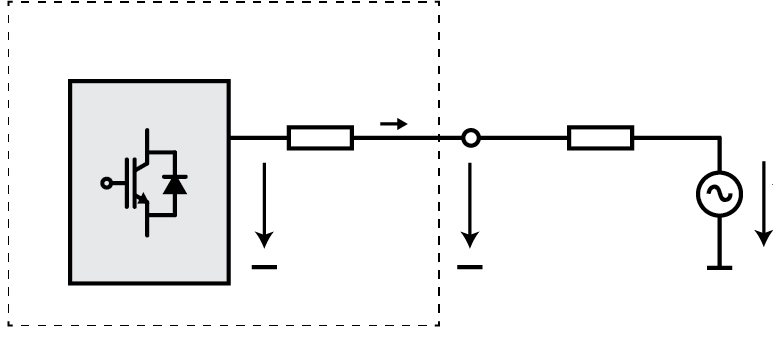}};
			\node[draw=none,fill=none] at (-.5,.55) {$L$};
			\node[draw=none,fill=none] at (1.7,.55) {$\cxZ_{\gr}(s)$};
			\node[draw=none,fill=none] at (-0.6,-.25) {$\cxv$};
			\node[draw=none,fill=none] at (0,0.575) {$\cxi$};
			\node[draw=none,fill=none] at (0.95,-.25) {$\cxE$};
			\node[draw=none,fill=none] at (3.05,-.35) {$\cxv_\gr$};
			\node[draw=none,fill=none] at (-1.45,1) {\scriptsize{Grid-forming converter}};
	\end{tikzpicture}}
	\caption{Example circuit of the converter under consideration.}
	\label{fig:system-model}
	
\end{figure}

\subsection{Converter system and current dynamics}

Figure~\ref{fig:system-model} depicts an example configuration with the circuit of the converter. 
On the converter-side of the PCC, we have the transformer inductance or an inductive input filter. The resistances herein are ignored, as their inclusion in the control scheme is trivial. This example configuration does not exclude the possibility of having an LCL filter or a shunt capacitor; the CL or the shunt capacitor could simply be considered part of the grid. The stiff voltage on the grid-side is illustrated here for the sake of presentation only. In this configuration, the source-side converter, which is not shown explicitly, is primarily responsible for controlling the DC-link voltages. Thus, to also simplify the analysis here, they are assumed to be realized by ideal voltage sources.\looseness=-1 

The current dynamics are shown in the (stiff) grid $\dq$-frame:
\begin{equation}\label{eq:pcc_dyn} 
	\cxv- j\omega_{1}L\cxi-	L\frac{d\cxi}{dt} = \cxE	\Leftrightarrow  (s+j\omega_{1})L\cxi = {\cxv - \cxE},
\end{equation}
where $\omega_1$ is the nominal synchronous frequency. The $\ab$-frame correspondence can be obtained by substituting $s \rightarrow s-j\omega_{1}$.

\subsection{UPSC with droop controllers}

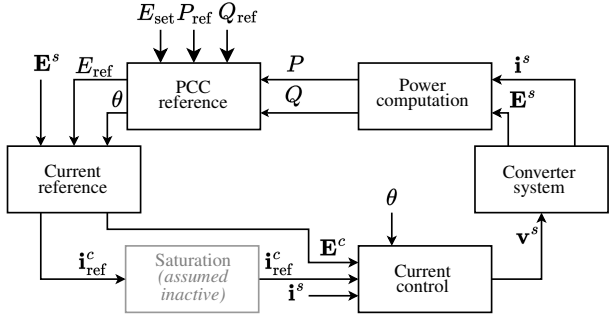
\begin{figure}[t]
	\centering
	{\begin{tikzpicture}[every text node part/.style={align=center}]
			\node[draw=none,fill=none] at (0,0){\includegraphics[trim={1.5cm 0.25cm 1cm 0.25cm},clip]{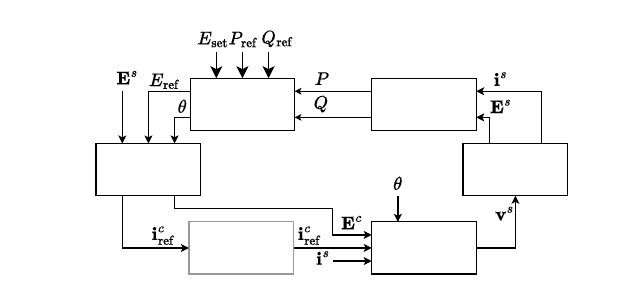}};;
			\node[draw=none,fill=none] at (1.55,0.8) {$\substack{\text{Power}\\ \text{computation}}$};
			\node[draw=none,fill=none] at (-3.1,-0.325) {$\substack{\text{Current}\\ \text{reference}}$};
			\node[draw=none,fill=none] at (-1.5,0.8) {$\substack{\text{PCC}\\ \text{reference}}$};
			\node[draw=none,fill=none] at (1.55,-1.65) {$\substack{\text{Current}\\ \text{control}}$};
			\node[draw=none,fill=none] at (-1.5,-1.65) {${\color{gray}\substack{\text{Saturation}\\ \textit{(assumed}\\ \textit{inactive)}}}$};
			\node[draw=none,fill=none] at (3.1,-0.35) {$\substack{\text{Converter}\\ \text{system}}$};
	\end{tikzpicture}}
	\caption{A high-level block diagram of the complete scheme.}
	\label{fig:control_scheme}
\end{figure}

The complete UPSC scheme is provided in Figure~\ref{fig:control_scheme}, and the definitions follow closely from~\cite{harnefors2020universal}. The complex power is $\cxS = \cxE\cxi^{*}= \cxE^s\cxi^{s*}$. The active power is $P=\re\{\cxS\}$, whereas the reactive power is $Q=\im\{\cxS\}$. The corresponding power references are denoted by $P_\rf$ and $Q_\rf$. The reference $E_\mathrm{set}$ is a PCC voltage magnitude setpoint. Remaining quantities are either defined later or are clear from the context.\looseness=-1

\subsubsection{PCC voltage reference generation}\hfill

\quad \textit{(a) Synchronization (angle):} The power synchronization loop generates the converter $\dq$-frame angle~$\theta$: 
\begin{equation}\label{eq:PSL}
	\theta = (1/s)\left[\omega_1 + K_p(s)(P_\rf - P)\right],
\end{equation}
where $K_p(s) = \frac{sT_d+1}{sM+k_m}$,
with $T_d$ being the time constant for damper winding emulation, $k_m$ being the droop constant, and $M$ being the inertia constant. \looseness=-1

\quad \textit{(b) Voltage magnitude reference:}
The PCC voltage magnitude reference is generated by
\begin{equation}\label{eq:droop}
	\begin{split}
		E_\rf = E_\ext &+ F_{Q}(s) \left[Q_\rf - H_{\alpha_Q}(s)Q\right]\\ &+ F_{P}(s) \left[P_\rf - H_{\alpha_P}(s) P\right],
	\end{split}
\end{equation}
where the first additive term is the $QV$ control, and the second additive term is the $PV$ control. 
\subsubsection{Current reference generation}
The current reference is given by the alternating voltage controller (AVC), designed as
{\medmuskip=.7432mu \thickmuskip=.7432mu \thinmuskip=.7432mu\begin{equation}\label{eq:psavc}
		\cxi_{\rf}^c= \frac{P_\rf-jQ_\rf}{E_\ext}+\frac{s+\alpha_{a}}{s(sL+R_{a})} \left(E_\rf-\cxE^c\right),
\end{equation}}where $R_a$ is the proportional gain of the inner current controller, $\alpha_a$ is the AVC integral control bandwidth, and $\cxE_\rf=E_\rf$.\footnote{In other words, the reference is aligned with the converter $\dq$-frame.} The first additive term is the reference feedforward proposed in~\cite{harnefors2020reference}. The saturation block is assumed to be inactive.\looseness=-1
\subsubsection{Current controller}
The current controller is also in the converter $\dq$-frame, while including a PCC voltage feedforward and accounting for the voltage drop across the inductance:
\begin{equation}\label{eq:cc}
\cxv^c_\rf = R_a(\cxi_\rf^c-\cxi^c)+j\omega_1L\cxi^c+H_{\alpha_F}(s)\cxE^c,
\end{equation}
where $R_a$ is the proportional gain, and $\cxv^c_\rf$ is the reference voltage (e.g., to be created via PWM), which is to be later mapped to the $\ab$-frame. Since we are interested in a lower frequency range behavior, we ignore the modulator and the related delays, and let $\cxv^c=\cxv^c_\rf$, see~\cite{harnefors2017vsc} for its inclusion. Moreover, it is straightforward to replace this part with different current
controllers (for example, linearized versions of model predictive current controllers such as~\cite{karaca2024damping,karaca2024frequency}) and repeat the analysis in subsequent sections.

\subsection{Preliminaries on passivity}
In the next section, the converter will be modeled with its Norton equivalent circuit, i.e., an ideal current source with a MIMO small-signal input admittance $\bm{Y}(s)$, i.e., $\Delta\bm{i}=-\bm{Y}(s)\Delta\bm{E}$. The zeros of the determinant of $I+\bm{Y}(s)\bm{Z}_{\gr}(s)$, which are the poles of the closed-loop system, can be used to characterize the small-signal stability, or one can directly apply the multivariable Nyquist criterion to the open-loop transfer matrix~\cite{maciejowski1989multivariable,harnefors2007modeling,sun2011impedance}.\looseness=-1

Alternatively, a passivity-oriented analysis allows assessing stability independent of the grid~\cite{willems1972dissipative,willems1972dissipative2}. If all elements connected to the grid are dissipative (i.e., passive) on their own, then the stability of the interconnected system is ensured. Such a condition holds for the grid itself if it consists only of resistive, inductive, or capacitive elements. On the other hand, a sufficient condition for dissipativeness of the converter system is given by the passivity index~\cite{harnefors2020asymmetric,bao1998new}.
Let $\lambda(\cdot)$ denote the operator that computes the set of eigenvalues of a given matrix. The passivity index at $\omega$ (with respect to the $\dq$-frame) is \looseness=-1
\begin{equation}
	\nu(\bm{Y}(s),\omega) = \frac{1}{2} \min\left\{ \lambda(\bm{Y}(j\omega)+\bm{Y}^*(j\omega)) \right\}.
\end{equation}
Then, the system is stable (and dissipative) if $\nu(\bm{Y}(s),\omega) >0,\ \forall \omega\geq0.$\footnote{Positive orthant restriction is valid due to $Y^*_{(\cdot)}(j\omega)=Y_{(\cdot)}(-j\omega)$.}
Pure passivity for the whole frequency range is impossible to obtain. Instead, stability can still be guaranteed by ensuring the condition holds for the range where resonances may appear~\cite{harnefors2007analysis}. In \cite{harnefors2007input}, it was suggested to set the lower limit close to 0.1$\,\mathrm{pu}$, considering the poorly damped resonances in typical grids~\cite{anderson1999subsynchronous}. Grid codes also generally do not impose input resistance requirements within a $\pm$0.05$\,\mathrm{pu}$ window around the nominal synchronous frequency. {As a remark, the literature considers it beneficial that the negative passivity index regions are as narrow as possible, however, one should exercise caution for conflicts of this objective with increasing the passivity index value~\cite{chen2023pitfalls}. Less conservative basis-change-based passivity certificates, as in~\cite{chen2024extended}, are considered out-of-scope.\looseness=-1

\section{Input Admittance of UPSC with Droop}
To obtain the small-signal input admittance form $\Delta\bm{i}=-\bm{Y}(s)\Delta\bm{E}$, three main steps are: defining the inner closed-loop system, incorporating the current reference, and finally bringing in the power synchronization and the droop control.
In the remainder, for any quantity $\cxx$, let $\cxx=\cxx_0+\Delta\cxx$, with $\cxx_0$ and $\Delta\cxx$ denoting the steady-state and small-signal quantities, respectively.
For the quantities with a reference, assume these references represent their steady-state values as well. We often transform between the converter $\dq$-frame and the grid $\dq$-frame, and this is given by ${\cxx}^c={\cxx}e^{-j\Delta\theta}\approx{\cxx}-j{\cxx}_0\Delta\theta$.\looseness=-1

\subsubsection{Inner closed-loop system}
Replacing the converter $\dq$-frame quantities (specifically, $\cxi^c$ and $\cxE^c$) in \eqref{eq:cc} with the grid $\dq$-frame ones, we obtain
{\begin{equation}\label{eq:converter_voltage1}
	\begin{split}
		\cxv^c = &R_a(\cxi_\rf^c-\cxi+j\cxi_{0}\Delta\theta)+j\omega_1L\cxi-j\omega_1Lj\cxi_{0}\Delta\theta\\
		&\quad+H_{\alpha_F}(s)\cxE-H_{\alpha_F}(s)jE_{\ext}\Delta\theta,
	\end{split}
\end{equation}}where $\cxi_{0}=\frac{P_\rf-jQ_\rf}{E_\ext}$. Mapping the converter voltage in \eqref{eq:converter_voltage1} to the grid $\dq$-frame gives
{\begin{equation}\label{eq:converter_voltage}
	\begin{split}
		\cxv = &\cxv^c + j(E_{\ext}+j\omega_1L\cxi_{0})\Delta\theta\\
		= &R_a(\cxi_\rf^c-\cxi+j\cxi_{0}\Delta\theta)+j\omega_1L\cxi\\
		&\quad+H_{\alpha_F}(s)\cxE-j[H_{\alpha_F}(s)-1]E_{\ext}\Delta\theta. \\
	\end{split}
\end{equation}}Plugging \eqref{eq:converter_voltage} into \eqref{eq:pcc_dyn}, and using the definitions $G_{c}(s)=\frac{R_a}{sL+R_a}$ and $Y_i(s)=\frac{H_{\alpha_F}(s)-1}{sL+R_a}$, we obtain the inner closed-loop system:
\begin{equation}\label{eq:inner_closed}
		 \cxi={G_{c}(s)}\cxi_\rf+{Y_i(s)}\cxE+j\left[G_{c}(s)\cxi_{0}-Y_i(s)E_{\ext}\right]\Delta\theta.                                                                                                                            
\end{equation}

\subsubsection{Including the current reference}
The UPSC current references in \eqref{eq:psavc} can be written as
{\begin{equation}\label{eq:psavc1}
\cxi_\rf^c=\cxi_{0}+{Y_{c}(s)}(E_{\rf}-\cxE+jE_{\ext}\Delta\theta),     
\end{equation}}where $Y_{c}(s)={\frac{s+\alpha_{a}}{s(sL+R_{a})}}$. Next, plugging \eqref{eq:psavc1} into \eqref{eq:inner_closed}, and using the definitions $Y_{c}'(s)=G_{c}(s)Y_{c}(s)$ and $Y_i'(s)=[Y_i(s)-G_{c}(s)Y_{c}(s)]$, we obtain
{\begin{equation}
	\begin{split}
		\cxi=&{G_{c}(s)}\cxi_{0}+{Y_{c}'(s)}E_{\rf} +{Y_i'(s)}\cxE\\
		&\quad+j\left[G_{c}(s)\cxi_{0}-Y_{i}'(s)E_{\ext}\right]\Delta\theta.\label{eq:mod_in}
	\end{split}                                                                                                                                                                            
\end{equation}}                                                                                  

\subsubsection{Including the PCC voltage reference}
Assuming constant $E_{\ext}$, $P_{\rf}$, $Q_{\rf}$,  we obtain small-signal models of \eqref{eq:PSL} and~\eqref{eq:mod_in}; combining them we get the following small-signal system
{\begin{equation}\label{eq:prelim}
		\begin{split}
		\Delta\cxi =& Y_{c}'(s)\Delta E_{\rf}+  Y_{i}'(s)\Delta \cxE \\
		&\quad+ j\left[G_{c}(s)\cxi_{0}-Y_{i}'(s)E_{\ext}\right]\frac{K_{p}(s)}{s}\Delta P.
	\end{split}
\end{equation}}Next, invoking the droop relations in~\eqref{eq:droop}, we get
	{\medmuskip=.01mu \thickmuskip=.01mu \thinmuskip=.01mu\begin{equation}
				\begin{split}
			\Delta\cxi = &Y_{i}'(s)\Delta \cxE\\
			&\hspace{0.05cm}-\Big\{ Y_{c}'(s){\tilde{F}_{P}(s)}  - j\left[G_{c}(s)\cxi_{0}-Y_{i}'(s)E_{\ext}\right]\frac{K_{p}(s)}{s}\Big\} \Delta P\\
			&-Y_{c}'(s){\tilde{F}_{Q}(s)}\Delta Q,
			\label{tog}
		\end{split}
	\end{equation}}where $\tilde{F}_{P}(s)=F_{P}(s)H_{\alpha_P}(s)$ and $\tilde{F}_{Q}(s)={F_{Q}(s)H_{\alpha_Q}(s)}$.

Notice that substituting $\Delta P$ and $\Delta Q$ from~\eqref{tog} yields the desired form, and completes the derivation. To this end, complex power can be linearized as
{\medmuskip=.7432mu \thickmuskip=.7432mu \thinmuskip=.7432mu\begin{equation}\begin{split}\cxS = \cxE\cxi^{*} &= (E_{\ext}+\Delta\cxE)(\cxi_{0}^{*}+\Delta\cxi^{*})\\ &\approx E_{\ext}\cxi_{0}^{*} + 
{(E_{\ext}\Delta\cxi^{*}+\cxi_{0}^{*}\Delta\cxE)}= \cxS_0+{\Delta\cxS}.\end{split}
\end{equation}}Thus, we have
	{\medmuskip=.01mu \thickmuskip=.01mu \thinmuskip=.01mu\begin{equation}\label{eq:powers}
		\begin{split}
\Delta P &= \re\{\Delta \cxS\}=(E_{\ext}\Delta i_{d}+i_{d0}\Delta E_{d}+i_{q0}\Delta E_{q}),\\
\Delta Q &= \im\{\Delta \cxS\}= (-E_{\ext}\Delta i_{q}-i_{q0}\Delta E_{d}+i_{d0}\Delta E_{q}).
		\end{split}
\end{equation}}Inserting \eqref{eq:powers} into \eqref{tog}, and switching from complex space vectors to real space vectors for all quantities, we get 
\begin{equation}
	 \bm{D}(s)\Delta\bm{i}=-\bm{W}(s)\Delta\bm{E},
\end{equation}
where  $\bm{D}(s)$ and $ \bm{W}(s)$ are provided in \eqref{eq:final} on top of the next page.
\begin{figure*}
{\medmuskip=.01mu \thickmuskip=.01mu \thinmuskip=.01mu\begin{equation}\label{eq:final}
	\begin{split}
		{\bm{D}(s)}&={\bmat{cc}
			1+Y_{c}'(s)\tilde{F}_{P}(s)E_{\ext}+G_{c}(s)i_{q0}\frac{K_{p}(s)}{s}E_{\ext} & -Y_{c}'(s)K_{Q}H_{Q}(s)E_{\ext} \\
			-\left[G_{c}(s)i_{d0}-Y_{i}'(s)E_{\ext}\right]\frac{K_{p}(s)}{s}E_{\ext} &1 \emat},	\\ 
		\bm{W}(s)&=\bmat{cc} 
			-{Y}_{i}'(s)+Y_{c}'(s)[\tilde{F}_{P}(s)i_{d0}-\tilde{F}_{Q}(s)i_{q0}]+G_{c}(s)i_{q0}\frac{K_{p}(s)}{s}i_{d0} & Y_{c}'(s)[\tilde{F}_{P}(s)i_{q0}+\tilde{F}_{Q}(s)i_{d0}]+G_{c}(s)i_{q0}^2\frac{K_{p}(s)}{s}
			\\	 
			-\left[G_{c}(s)i_{d0}-Y_{i}'(s)E_{\ext}\right]\frac{K_{p}(s)}{s}i_{d0} & -{Y}_{i}'(s)-\left[G_{c}(s)i_{d0}-Y_{i}'(s)E_{\ext}\right]\frac{K_{p}(s)}{s}i_{q0}
			\emat.
	\end{split}
\end{equation}}\vspace{0.1cm}

\hrulefill
\end{figure*}
Finally, the input admittance of the converter system governed by
UPSC with droop controllers is given by $\bm{Y}(s)=\bm{D}^{-1}(s)\bm{W}(s).$
\section{Sensitivity Analysis for Passivity Index}

\begin{table}[t!]
	\caption{The base case}
	\label{table:base}
	\vspace{.1cm}
	\centering
	\begin{tabular}{lc|lc}
		\hline
		Parameter & Value & Parameter & Value \\
		\hline
		$k_m$ &  $\SI{20}{{\pu}}$ & $K_Q$ & $\SI{0.1}{{\pu}}$  \\ 
		$T_d$ & $\SI{15}{{\pu}}$ & $\alpha_Q$ & $\SI{0.5}{{\pu}}$ \\ 
		$M\,(\text{Time constant:}H)$ &  $\SI{565}{\pu}\,(\SI{0.75}{s}) $ & 	$L$ &  $\SI{0.15}{{\pu}}$ \\
		$K_P$ & $\SI{0.1}{{\pu}}$ & $R_a$ &  $\SI{0.3}{{\pu}}$ \\
		$K_{P,I}$ & $\SI{0}{{\pu}}$ 	& $\alpha_a$ &  $\SI{0.025}{{\pu}}$\\ 
		$\alpha_P$ & $\SI{0.5}{{\pu}}$  & $\alpha_F$ & $\SI{2}{{\pu}}$ \\
		\hline
	\end{tabular}
\end{table}

In this section, we provide a sensitivity analysis for the passivity index in the low-frequency range, specifically, up to  $0.20\,\mathrm{pu}$, i.e., $12$ Hz in a $60$ Hz grid.
The base case is provided in Table~\ref{table:base} for the voltage controllers  ${F}_{Q}(s)=K_Q$ and ${F}_{P}(s)=K_P+\frac{K_{P,I}}{s}$. Here, a PI controller is chosen for $PV$ control as it is considered essential in diode rectifier applications~\cite{yu2019analysis,karaca2025effective}. Sensitivity analyses of other variants are also straightforward using the input admittance derived in the previous section.

Figure~\ref{fig:QV} illustrates the passivity index when $K_Q$ is varied for the cases of $(P_\rf,Q_\rf)=(0,0)$ and $(P_\rf,Q_\rf)=(1,0.5)$. Increasing $K_Q$ boosts the passivity index in the very low-frequency range. However, further increases lead to a significant drop at higher frequencies, resulting in a much higher zero-crossing frequency for the passivity index.\looseness=-1

\begin{figure}[t]
	\begin{subfigure}{0.495\columnwidth}
		\centering
	\includegraphics[width=1\columnwidth]{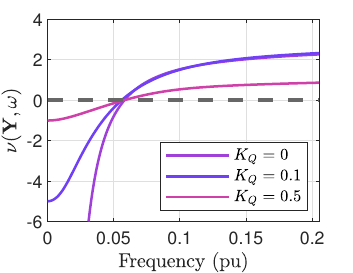}
	\caption{$(P_\rf,Q_\rf)=(0,0)$.}
	\end{subfigure}
	\begin{subfigure}{0.495\columnwidth}
		\centering
		\includegraphics[width=1\columnwidth]{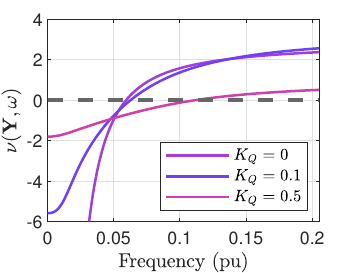}
		\caption{$(P_\rf,Q_\rf)=(1,0.5)$.}
	\end{subfigure}
	\caption{Sensitivity of $\nu(\bm{Y}(s),\omega)$ to $K_Q$.}\label{fig:QV}
\end{figure}

Figure~\ref{fig:PV} illustrates the passivity index when $K_P$ is varied for the cases of $(P_\rf,Q_\rf)=(0,0)$ and $(P_\rf,Q_\rf)=(1,0.5)$. Increasing $K_P$ boosts the passivity index without posing any risk of an increase in the zero-crossing frequency.

\begin{figure}[t]
	\begin{subfigure}{0.495\columnwidth}
		\centering
		\includegraphics[width=1\columnwidth]{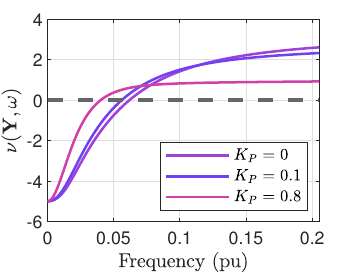}
		\caption{$(P_\rf,Q_\rf)=(0,0)$.}
	\end{subfigure}
	\begin{subfigure}{0.495\columnwidth}
		\centering
		\includegraphics[width=1\columnwidth]{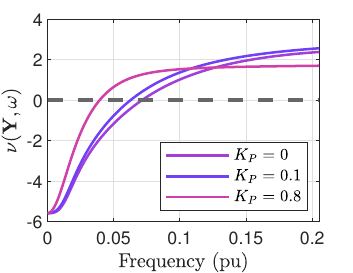}
		\caption{$(P_\rf,Q_\rf)=(1,0.5)$.}
	\end{subfigure}
	\caption{Sensitivity of $\nu(\bm{Y}(s),\omega)$ to $K_P$.}\label{fig:PV}
\end{figure}

Figure~\ref{fig:PVI} illustrates the passivity index when $K_{P,I}$ is varied for the cases of $(P_\rf,Q_\rf)=(0,0)$ and $(P_\rf,Q_\rf)=(1,0.5)$. Increasing $K_{P,I}$ creates a larger dip in the passivity index. As can be seen in $(P_\rf,Q_\rf)=(0,0)$, this does not bring in any benefits in lowering the zero-crossing frequency for the passivity index.
As aforementioned, an integral control gain in $PV$ control could be considered essential in diode rectifier applications~\cite{yu2019analysis,karaca2025effective}, and this observation could imply a control performance limitation in ramping up active power over a diode rectifier. Though not explicitly provided due to page limits, the following changes have a negative effect on the passivity index in the low-frequency range: increasing $\alpha_a$ or $T_d$, decreasing $M$ or $R_a$.

\begin{figure}[t]
	\begin{subfigure}{0.495\columnwidth}
		\centering
		\includegraphics[width=1\columnwidth]{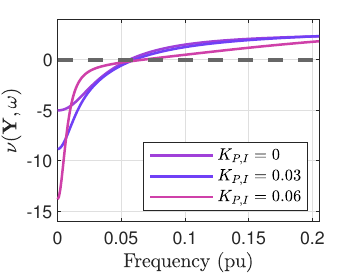}
		\caption{$(P_\rf,Q_\rf)=(0,0)$.}
	\end{subfigure}
	\begin{subfigure}{0.495\columnwidth}
		\centering
		\includegraphics[width=1\columnwidth]{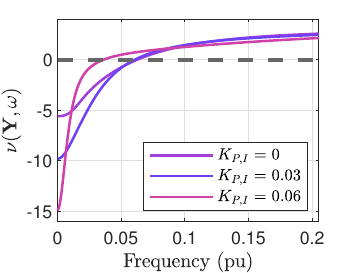}
		\caption{$(P_\rf,Q_\rf)=(1,0.5)$.}
	\end{subfigure}
	\caption{Sensitivity of $\nu(\bm{Y}(s),\omega)$ to $K_{P,I}$.}\label{fig:PVI}
\end{figure}

\section{Case Study}
\begin{figure}[t!]
	\centering
	{\begin{tikzpicture}[every text node part/.style={align=center}]
			\node[draw=none,fill=none] at (0,0){\includegraphics[scale=1.3]{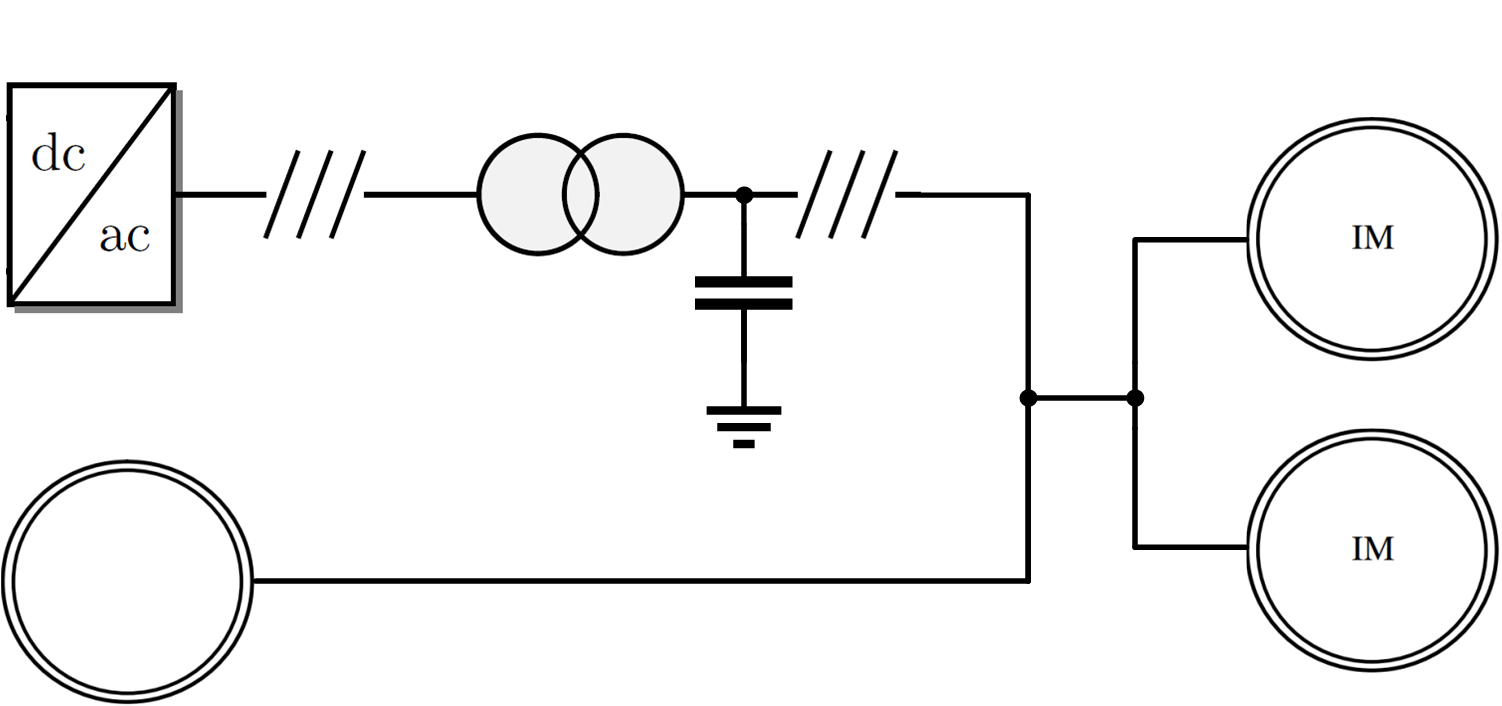}};
			\node[draw=none,fill=none] at (-3.95,.7) {$\substack{\text{Grid-forming}\\ \text{converter}}$};
			\node[draw=none,fill=none] at (-3.95,-1) {$\substack{\text{Synchronous}\\ \text{generator}}$};
			\node[draw=none,fill=none] at (-2.665,-0.985) {\tiny SG};
	\end{tikzpicture}}
	\caption{Simulation environment.}
	\label{fig:sys_sfc}
\end{figure}
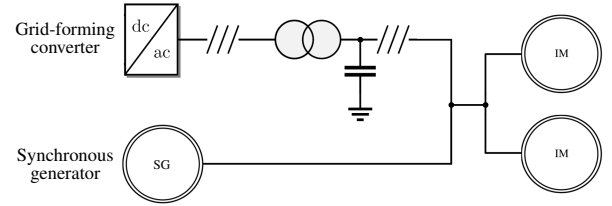

We conduct a study in an environment implemented in MATLAB Simulink. The setup, illustrated in Figure~\ref{fig:sys_sfc}, consists of a grid-forming converter with a shunt capacitor operated in parallel with a synchronous generator feeding two identical induction motors. The relevant system parameters and rated values are given in Table~\ref{tab:rated values}. Other parameters and control loops related to the synchronous generator and induction motors are considered confidential. Additionally, the objective of passivity is to assess the stability of the converter independent of the rest of the grid.

The grid-forming converter implements the UPSC scheme shown in Figure~\ref{fig:control_scheme}, with parameters listed in Table~\ref{table:base}. Since our focus is on system behavior in the low-frequency range, we work with averaged converter models. However, the switching and control system delays are taken into account, e.g., by considering the actual sampling intervals and control time scales of our converters. No power references are provided to demonstrate the power sharing capabilities. To abide by the page limitations, we present only the essential plots in the figures. 

A torque ramp from the induction motors begins at $\SI{7}{{s}}$. Once steady-state is reached, at $\SI{10}{{s}}$, several control parameters of only the grid-forming converter are adjusted from the base case. Specifically, the parameters are modified as follows: $K_P=\SI{0.05}{{\pu}}$, $K_Q=\SI{0.05}{{\pu}}$, and $\alpha_a=\SI{0.075}{{\pu}}$.\footnote{This modification implies also a shift in reactive power sharing. Moreover, it may not be possible to freely change the droop gains, as they might be dictated by the grid operator.} The resulting change in passivity index at this steady-state operating point is shown in Figure~\ref{fig:caset}. The active and reactive power plots are provided in Figure~\ref{fig:cases}. The PCC voltage measurements and references for the grid-forming converter are provided in Figure~\ref{fig:cases2}. A power oscillation between the grid-forming converter and the synchronous generator, at approximately $\SI{6}{{Hz}}$, can be observed starting from $\SI{10}{{s}}$. This is consistent with the reduction in passivity index at that frequency, as shown in Figure~\ref{fig:caset}. To restore the system stability,  the control parameters are reverted to their original values at $\SI{10.4}{s}$. 

\begin{table}[t!]
	\caption{System parameters and rated values.}
	\label{tab:rated values}
		\vspace{.1cm}
	\centering
	\begin{tabular}{lcl}
		\hline
		Parameter & Symbol & Value \\
		\hline
		Rated power of the transformer/converter & $S_{\mathrm{R},\mathrm{trafo}}$ & $\SI{18}{{\mega\voltampere}}$ \\ 
		Rated power of the generator & $S_{\mathrm{R},\mathrm{SG}}$ & $\SI{18}{{\mega\voltampere}}$ \\
		Rated power of the induction motor & $S_{\mathrm{R},\mathrm{IM}}$ & $\SI{5}{{\mega\voltampere}}$ \\
		Rated voltage at the primary (grid) & $V_{\mathrm{R},\mathrm{prim}}$ & $\SI{13.8}{\kilo\volt}$ \\
		Rated voltage at the secondary (converter) & $V_{\mathrm{R},\mathrm{sec}}$ & $\SI{3.0}{\kilo\volt}$ \\
		Rated angular frequency &  $\omega_1$ & $\SI[parse-numbers=false]{2 \pi 60}{\radian\per\second}$ \\
		\hline
	\end{tabular}
\end{table}

\begin{figure}[t]
	\centering
	\begin{subfigure}{0.575\columnwidth}
		\centering
		\includegraphics[width=1\columnwidth]{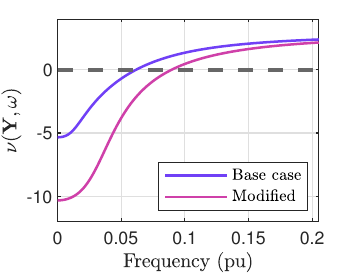}
	\end{subfigure}
	\caption{Passivity index of the converter.}\label{fig:caset}
\end{figure}

\begin{figure}[t]
	\centering
	\begin{subfigure}{0.999\columnwidth}
		\centering
		\includegraphics[width=1\columnwidth]{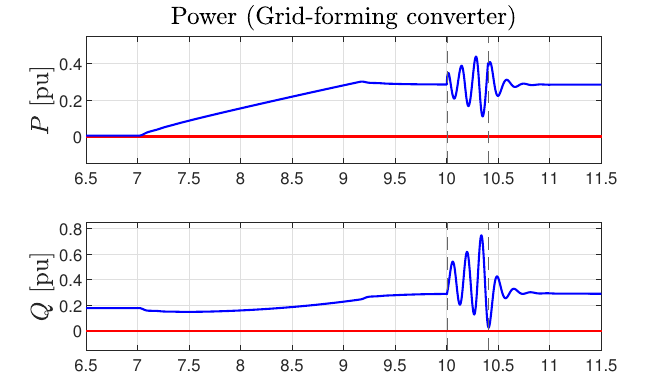}
	\end{subfigure}
	\centering
	\begin{subfigure}{0.999\columnwidth}
		\centering
		\includegraphics[width=1\columnwidth]{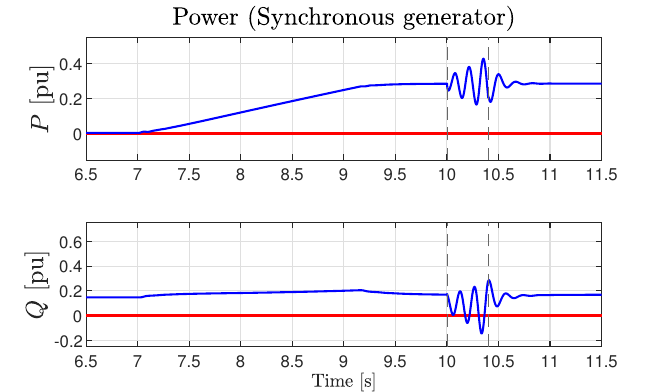}
	\end{subfigure}
	\caption{Active and reactive power measurements and references. \textit{(Red: Reference. Blue: Measurement.)}}\label{fig:cases}
\end{figure} 
\begin{figure}[t]
	\centering
		\includegraphics[width=1\columnwidth]{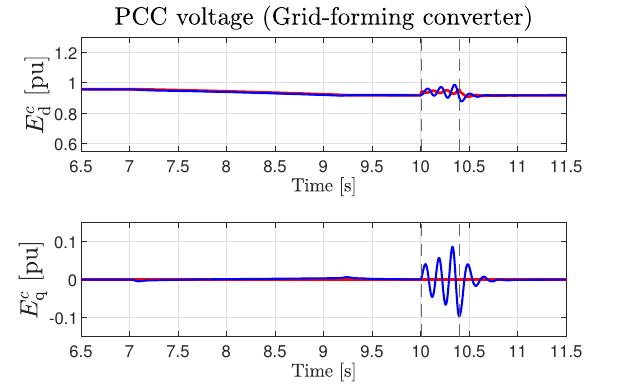}
	\caption{Converter $\dq$-frame PCC voltage measurements and their references generated by the $QV$ and $PV$ controllers. \textit{(Red: Reference. Blue: Measurement.)}}\label{fig:cases2}
\end{figure} 

\section{Conclusion}
We derived the input admittance of UPSC with $QV$ and $PV$ droop controllers. The passivity index resulting from this input admittance illustrated the benefits of the proportional components of these controllers. 
Conversely, a large integral control gain in $PV$ control led to a passivity dip, which could be an important consideration in certain applications. A case study demonstrated that the simulations confirm the insights obtained from the passivity index when droop gains are reduced and the AVC integral bandwidth is increased. 
This case study also confirms that, for the control scheme under consideration, the passivity index serves as a meaningful indicator for a stand-alone stability assessment independent of the rest of the grid. \looseness=-1

\bibliographystyle{IEEEtran}
\bibliography{virtual_report}

\end{document}